\documentclass[11pt,twoside]{article}

\usepackage{asp2006}
\usepackage{epsf}
\usepackage{epsfig}
\usepackage{lscape}

\begin{document}
\title{Studying the Structure of the Stellar Wind in LS 5039}    
\author{V. Bosch-Ramon}    
\affil{Max Planck Institut f\"ur Kernphysik, Saupfercheckweg 1, Heidelberg 69117, Germany; 
vbosch@mpi-hd.mpg.de}    

\begin{abstract} 
The binary LS 5039 is a non-thermal X-ray emitter that presents jet-like radio structures, and is also one of the most
misterious TeV sources in our Galaxy. The presence of an O-type star in LS 5039 implies that the non-thermal emitter must be
embedded in a strong stellar wind, and the role of the latter could be relevant for the understanding of the high-energy
behavior of the source. In this work, we show that the lack of absorption features in the soft X-ray spectrum of LS 5039 can
constrain strongly the parameters that describe the wind, and ultimately the location of the non-thermal emitter.
\end{abstract}

\section{Introduction}  

The X-ray binary LS~5039 presents non-thermal radio and very high-energy emission (e.g. Mart\'i et al. 1998; Aharonian et al.
2005), and the X-rays emitted by this object are also of likely non-thermal origin (e.g. Bosch-Ramon et al. 2007 -B07-; see
also Takahashi et al. 2009). The system consists of an O main sequence star and a compact object of unknown nature (Casares
et al. 2005 -C05-), and presents jet-like or collimated radio structures (e.g. Paredes et al. 2000, 2002; Rib\'o et al.
2008). 

LS~5039 was observed in 2005 by {\it XMM-Newton} right after the periastron passage (phase $\simeq 0.04$; OBS1), close to the
superior conjunction of the compact object (SUPC=0.06), and around the apastron passage (phase $\simeq 0.5$; OBS2), slightly
before the inferior conjunction (INFC=0.72)\footnote{New ephemeries have been presented by Aragona et al. (2009), although
they are similar to those of C05 and do not change the conclusions of this work.}. These observations, presented in detail in
B07, showed that photo-electric absorption may only take place in the interstellar medium, being absorption in the stellar
wind negligible, and put an upper-limit of $10^{21}$~cm$^{-2}$ on the difference between the absolute hydrogen column
densities of OBS1 and OBS2 ($\Delta N_{\rm H}$). This upper limit is about an order of magnitude smaller than the value
predicted adopting a homogeneous spherical stellar wind (B07). Possible explanations put forward in B07 for such a small
photo-electric absorption were a strongly ionized wind, which seems unlikely due to the moderate X-ray luminosity of the
source of few times $10^{33}$~erg~s$^{-1}$, an X-ray emitter far from the compact object at the borders of the binary system,
and finally an inhomogeneous wind with a large porosity length (see, e.g., Owocki \& Cohen 2006). Here, we explore the
possibility that the porosity length is long enough to effectively reduce $N_{\rm H}$ at large, and investigate what are the
implications of this for the stellar wind.

\section{Constraints on the stellar wind properties from soft X-ray data} 

The constraints on the $N_{\rm H}$ inferred by B07 can be used to derive restrictions on the properties of the stellar wind
of LS~5039. For this, we adopt a simple clumpy spherical wind model in which the clumps are assumed to be spherical with a
constant radius of $R_{\rm c}=10^{10}\,R_{\rm c~10}$~cm. The wind volume filling factor is fixed to its value
$f_0=dV'_0/dV_0=0.01\,f_{0.01}$ at the periastron distance $a=r_0\sim 1.4\times 10^{12}$~cm from the star. Here, $dV'_0$ and
$dV_0$ are the volume occupied by clumps and total one, respectively, for a {\it shell} located at $r_0$. At the same
distance, the average wind mass density is $\rho_0=9\times 10^{-15}$~cm$^{-3}$, as inferred from the observed $H_{\alpha}$ EW
assuming an homogeneous and isotropic wind (C05). The correction for clumpiness would require $\rho_0\,f_0^{1/2}$ (see,
e.g., Owocki \& Cohen 2006). The wind/clump speed is fixed to its value at $r_0$, i.e. $v_0=1.5\times 10^8$~cm~s$^{-1}$.

Here, we are interested on finding the $f_0$ dependence of the $\Delta N_{\rm H}$, and the $f_0$ and $R_{\rm c}$ dependences
of the number of clumps in the line of sight ($N_{\rm c}$) to derive, using $\Delta N_{\rm H}$, restrictions on the wind
parameters. $N_{\rm c}$ is relevant given the fact that, although a single clump could be optically thick, the probability to
catch a clump may be significantly smaller than 1. In that case, since the {\it XMM-Newton} exposure was similar or smaller
to the time needed by a clump to cover the interclump typical distance $\sim R_{\rm c}/f_0$, we can assume that the duration
of these observations is not relevant.

For the simplicity of the employed formulas, we consider the system/X-ray emitter/observer geometry for OBS1 and OBS2 to be
the same as the one for SUPC and INFC. Given the moderate eccentricity of the system, this simplification should not affect
much the final results. We also neglect $N_{\rm H}$ around INFC when compared to the value at SUPC, i.e. $\Delta N_{\rm
H}\sim N_{\rm H~SUPC}$. 

From the model and observational constraints presented above, we can obtain the formulae for $N_{\rm c}$ and $\Delta
N_{\rm H}$ (around SUPC). For $N_{\rm c}$, we derive first:
\begin{equation}
K=\frac{3\,r_0^2f_0^{3/2}(1-\epsilon^2)}{4\,R_{\rm c}a} =
\frac{3\,r_0f_0^{3/2}(1-\epsilon^2)}{4Rc}=0.1\,\left(\frac{f_{0.01}^{3/2}}{Rc10}\right)\,,
\end{equation}
where $\epsilon$ is the system eccentricity $\epsilon=0.35$ (C05), and then
\begin{equation}
N_{\rm c}=\frac{K\,(\pi-\beta)}{\sin\beta(1-\epsilon cos\phi)}\approx 
(0.4-0.8)\,\left(\frac{f_{0.01}^{3/2}}{R_{\rm c~10}}\right)\,,		
\end{equation}
where $\phi$ is the orbital angle from periastron, which is small in our case, 
and $\beta$ the angle between the line of sight and the
line joining the compact object and the star, and $\beta=30^{\circ}$ and $60^{\circ}$, which correspond to inclination
angles of $i=60^{\circ}$ (neutron star case) and $30^{\circ}$ (black hole case; see C05). From $N_{\rm H}$, we can obtain 
$\Delta N_{\rm H}$:
\begin{equation}
\Delta N_{\rm H}=\frac{N_{\rm c}R_{\rm c}\rho_0}{f_0 m_{\rm H}}\approx (2-4)\times 10^{21}\,f_{0.01}^{1/2}\,{\rm cm}^{-2}\,.				
\label{eq3}
\end{equation}
The free parameters of our model are the inclination angle $i$ and the wind filling factor $f_0$.

\begin{figure*}[]
\begin{center}
\includegraphics[width=0.55\textwidth]{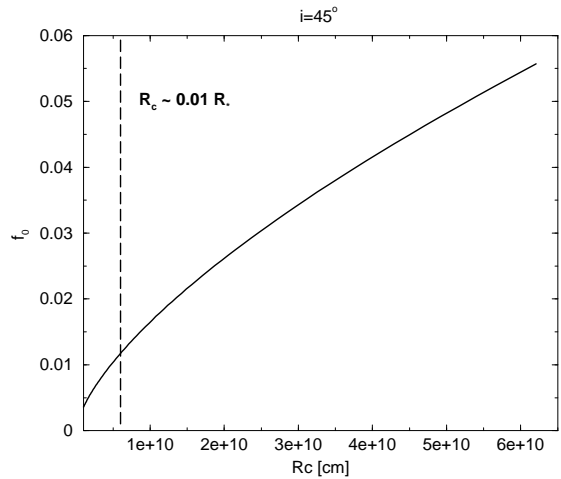}
\caption{The filling factor of the wind as a function of the clump size, derived assuming an X-ray emitter located  close to
the compact object and $i=45^{\circ}$, is shown. We note that for reasonable values of $R_{\rm c}$,  $f_0$ is about 0.01,
which  would imply a mass-loss rate 10 times smaller than that inferred for an homogeneous wind. In case $R_{\rm c}\la
109$~cm,  $f_0$  would not depend anymore on $R_{\rm c}$, and should be $\le 0.001$.}
\label{plots1}
\end{center}
\end{figure*}

\begin{figure*}[]
\begin{center}
\includegraphics[width=0.55\textwidth]{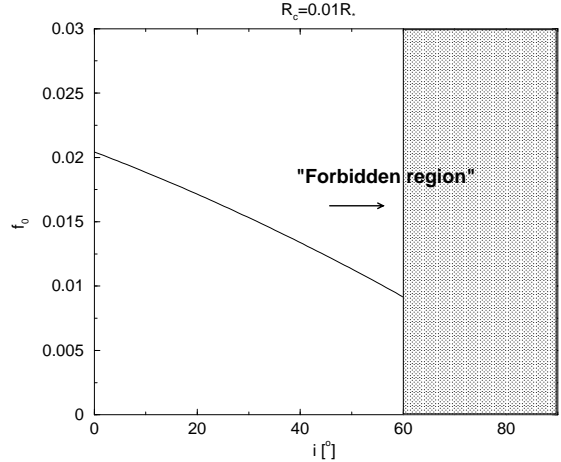}
\caption{The filling factor as a function of the inclination angle for a clump size of $0.01\,R_*$. The inclination 
dependence is rather smooth, and $f_0$ becomes smaller when $i$ grows. We note that for $i>60^{\circ}$ the X-ray would  pass
very close from the stellar surface around periastron/SUPC, implying that $f_0$ is not meaningful anymore. Actually,  for
such a $i$-values, the hydrogen column density would be much larger than the one assumed here, as discussed in  the text. In
such a case, soft X-rays should be absorbed even for small $\dot{M}_{\rm w}$, and they are not.}
\label{plots2}
\end{center}
\end{figure*}

We note that observations impose $\Delta N_{\rm H}<10^{21}$~cm$^{-2}$. This restriction plus the formulae shown above
actually imply that, either $N_{\rm c}<1$, and then $f_0\la (0.01-0.02)\,R_{\rm c~10}^{2/3}$, or $N_{\rm c}>1$, and then
$f_0\la 0.0004-0.003$. For $N_{\rm c}<1$ and a reasonable value of $R_{\rm c}$, e.g. $R_{\rm c}=0.01\,R_*=6.5\times 10^9$~cm
($R_*\simeq 6.5\times 10^{11}$~cm; C05), the wind filling factor should be $f_0\la 0.01$. In Fig.~\ref{plots1}, we plot $f_0$
as a function of the $R_{\rm c}$ for $i=45^{\circ}$ under $N_{\rm c}<1$, the most stringent condition as long as $R_{\rm
c}\ge 10^9$. We also plot in Fig.~\ref{plots2} the $i$-dependence of $f_0$ for $R_c=0.01\,R_*$ ($N_c < 1$).

\section{Discussion}

There are several caveats concerning the adopted wind model. The first one is the fact that for $i\ga 60^{\circ}$ the light
rays during SUPC should have a quite small impact parameter with respect to the star. It would imply that the wind structure
should be smoother, the density higher and the velocity lower. This would increase $\Delta N_{\rm H}$ leading to stronger
constraints on the real stellar mass-loss rate than those derived above. Also, even if we are not in a so extreme case as
that of $i=60^{\circ}$, the progressive wind acceleration means a lower wind speed at distances from the star shorter than
$r_0$, again effectively increasing the $\Delta N_{\rm H}$. Finally, the clump could have time to expand significantly from
the region it was formed. Accounting for a clump sound speed $\sim 10^6$~cm~s$^{-1}$ ($T\sim 10^4$~K) and an initial size
$R_{\rm c~0}<10^{10}$~cm, its size would grow significantly, i.e. $\Delta R_{\rm c}>R_{\rm c~0}$, increasing the interaction
probability and eventually entering in the regime described by Eq.~(\ref{eq3}) with the original $f_0$, since it depends only
on $\rho_0\,f_0^{1/2}$. 

It is worth noting that a wind suffering strong ionization would yield a lower $\Delta N_{\rm H}$ than the real value,
provided that there are no atoms to be ionized by X-rays. It is certainly happening in bright X-ray sources like Cygnus~X-1,
but LS~5039 is a relatively faint X-ray source. The X-ray luminosity around SUPC is few times $10^{33}$~erg~s$^{-1}$,
yielding a wind ionization region much smaller than the binary system size (B07). 

It could also happen that we are underestimating the real wind absorption with our X-ray fits if there were an additional
soft X-ray component on the top of the power-law one. This possibility cannot be discarded, and may indicate the presence of
a so far undetected accretion disk. Otherwise, the soft X-ray spectrum may be harder, and then we would be overestimating the
$\Delta N_{\rm H}$. In such a case, the constraints on the wind clumpiness and mass-loss rate would be even stronger.

We remark that one can easily derive constraints for the wind properties when the emitter is not located right on the compact
object but at different locations in relation to the star. For that, the angle dependence with $\beta$ is the same, but one
must modify accordingly the $\beta-i$ relationship. In addition, the distance from the emitter to the star center is not the
orbital distance anymore, and should be replaced by the proper distance to derive Eqs.~1, 2 and 3. This allows one for
instance to find at which distance the X-ray emitter should be from the star in case the wind were (approximately)
homogeneous. Interestingly, in such a case, as already noted by B07, the X-ray emitter should be located at a distance larger
than the orbital radius from the star and also from the compact object. A more detailed study will be presented elsewhere.

\subsection{Consequences of a deep emitter}

Putting aside the possibility that the X-ray emitter is far from the compact object, i.e. $\ga a$ from it, there remains only
a very porous and therefore weak wind as an alternative. The value of the mass-loss rate from the $f_0$ values derived above
is $\dot{M}_{\rm w}\sim 5\times 10^{-8}$~M$_{\odot}$/yr. This mass-loss rate implies a very weak wind momenta and, if the
compact object in the system were a non-accreting pulsar (see Martocchia et al. 2005; Dubus 2006; Sierpowska-Bartosik \&
Torres ), the spin-down luminosity of the pulsar wind required to explain the observations, of several
$10^{36}$~erg~s$^{-1}$, would generate a shock with a fairly wide opening angle, yielding an uncollimated structure
(Bogovalov et al. 2008, Romero et al. 2007)\footnote{We note however that opening angles close to $\pi$ would imply an X-ray
emitter of a size similar to that of the binary system, making our calculations not applicable.}. This would also mean that
the radio pulses would not be free-free absorbed during a substantial fraction of the orbit. All this would imply that, in
the pulsar scenario, the shocked pulsar outflow could hardly be the origin of the collimated radio emission, and pulses could
be detectable. On the other hand, we also note that such a weak stellar wind ($f_0\sim 0.01$) would be hardly enough to power
the observed non-thermal emission via accretion. We do not investigate here neither how a very porous stellar wind can
interact with the pulsar wind leading to non-thermal emission, nor how a very clumpy wind can lead to accretion and jet
formation, but these issues should be also addressed. 

Given the mentioned problems in the two proposed scenarios, a far X-ray emitter would be preferable, which is not compatible
with an X-ray emitter between the star and the pulsar, and would favor an emitter based on some sort of jet-like structure,
possibly like the radio and the TeV emission (see the discussion in Bosch-Ramon 2009; Khangulyan et al. 2008, Bosch-Ramon et
al. 2008), regardless the nature of the compact object, as noted in B07.

\section{Summary}

In this work, we show that the lack of significant features due to stellar wind absorption in the X-ray spectrum of LS~5039,
assuming an emitter close to the compact object, indicates that the wind has a filling factor $\sim 0.01$ or smaller. Such a
low wind filling factor would imply a stellar mass-loss rate one order of magnitude below the values inferred by C05, which
would have several implications in both the pulsar and the accretion scenario. In the former, such a weak wind would imply a
very large opening angle for the shocked wind structures. In such a case, the lack of radio pulses could not be explained by
free-free absorption in the stellar wind for a significant part of the orbit, nor the shocked pulsar wind would appear
collimated unless $i$ were close to $0^{\circ}$ or $90^{\circ}$. In the accretion scenario, such a low mass-loss rate could
hardly sustain the power of the observed persistent non-thermal bradband emission. 

An emitter located far from the compact object would not have the problems mentioned above for the two proposed scenarios,
and the emitter would be consistent with a jet-like emitter.

\acknowledgements 
W.B-R. thanks Anabella T. Araudo, Dmitry Khangulyan, Stan Owocki and Gustavo E. Romero for fruitful discussion on the topic
of this work. V.B-R. gratefully acknowledges support from the Alexander von Humboldt Foundation. V.B-R. acknowledges support
by DGI of MEC under grant AYA2007-68034-C03-01, as well as partial support by the European Regional Development Fund
(ERDF/FEDER).

\end{document}